\documentclass[a4paper,fleqn,usenatbib]{mnras}

\usepackage[T1]{fontenc}
\usepackage{ae,aecompl}

\usepackage{graphicx}	
\usepackage{amsmath}	
\usepackage{amssymb}	

\title[Magnetic activity on SZ~Psc]
        {Prominence activation, optical flare, and post-flare loops on the RS Canum Venaticorum star SZ Piscium}
\author[D. Cao et al.]
            {Dongtao Cao,$^{1, 2}$\thanks{E-mail: dtcao@ynao.ac.cn}
             Shenghong Gu,$^{1, 2, 3}$\thanks{E-mail: shenghonggu@ynao.ac.cn}
             Jian Ge,$^{4}$
             Tinggui Wang,$^{5}$
             Jilin Zhou,$^{6}$
\newauthor 
             Liang Chang,$^{1,2}$
             U. Wolter,$^{7}$
             M. Mittag,$^{7}$
             J. H. M. M. Schmitt,$^{7}$
             and V. Perdelwitz$^{7}$
\\
$^{1}$Yunnan Observatories, Chinese Academy of Sciences, Kunming 650216, China\\
$^{2}$Key Laboratory for the Structure and Evolution of Celestial Objects, Chinese Academy of Sciences, Kunming 650216, China\\
$^{3}$University of Chinese Academy of Sciences, Beijing 100049, China\\
$^{4}$Department of Astronomy, University of Florida, Bryant Space Science Center, Gainesville, Florida~32611, USA\\
$^{5}$Key Laboratory for Research in Galaxies and Cosmology, Department of Astronomy, University of Science and Technology\\ 
of China, Chinese Academy of Sciences, Hefei 230026, China\\
$^{6}$Department of Astronomy and Key Laboratory of Modern Astronomy and Astrophysics in Ministry of Education, Nanjing\\
University, Nanjing 210093, China\\
$^{7}$Hamburger Sternwarte, Universit\"{a}t Hamburg, Hamburg 21029, Germany
}

\date{Accepted 2018 October 10. Received 2018 October 9; in original form 2018 June 6}

\pubyear{2018}

\begin{document}
\label{firstpage}
\pagerange{\pageref{firstpage}--\pageref{lastpage}}
\maketitle

\begin{abstract}
We present the results of time-resolved high-resolution spectroscopic observations of the very active RS Canum Venaticorum (RS~CVn) star SZ~Piscium (SZ~Psc), obtained during 
two consecutive observing nights on October 24 and 25, 2011. Several optical chromospheric activity indicators are analyzed using the spectral subtraction technique, which show the 
remarkably different behavior between two nights. Gradually blue-shifted and strengthened excess absorption features presented in the series of the subtracted spectra (especially 
for the H$_{\alpha}$, $\mbox{He~{\sc i}}$ D$_{3}$ and H$_{\beta}$ lines), as a result of active stellar prominence that is rising its height along the line of our sight, was detected 
in the observations on October~24. This prominence activation event was probably associated with the subsequently occurred optical flare, and part of that flare decay phase was 
hunted in the observations on October~25. The flare was characterized by the prominent $\mbox{He~{\sc i}}$ D$_{3}$ line emission, as well as stronger chromospheric emission in 
the H$_{\alpha}$, H$_{\beta}$ and other active lines. The gradual decay of flare was accompanied by an obviously developmental absorption feature in the blue wing of the 
H$_{\alpha}$ and other active lines, which could be explained as cool post-flare loops which projected against the bright flare background. Therefore, a series of possibly associated 
magnetic activity phenomena, including flare-related prominence activation, optical flare and post-flare loops, were detected during our observations. 
\end{abstract}

\section{Introduction}
\indent
Solar-type activity phenomena including starspots, plages, flares and prominences, have been widely observed in many cool stars \citep{Schrijver2000}. It is commonly assumed that all 
of these active phenomena arise from a powerful magnetic dynamo generated by the interplay between the turbulent motions in the convection zone and the stellar differential rotation, 
in a manner similar to the Sun. Therefore, the solar activity paradigm provides a good reference to investigate magnetic activity phenomena encountered in solar-like stars. To detect stellar 
prominences and capture the other related transient activity events, in our present work, we performed time-resolved spectroscopic observations for SZ~Psc.\\
\indent
The star SZ~Psc (HD~219113, SAO~128041) is a well known double-lined spectroscopic and partial eclipsing binary system with an orbital period of about 3.97~days. The system consists 
of a more massive K1~IV primary component which fills 80--90\% of its Roche lobe and a F8~V secondary companion \citep{Jakate1976}. Based on the star's changing systemic velocity, 
\citet{Eaton2007} argued that there is a tertiary component in the system. Furthermore, applying the least squares deconvolution (LSD) method, \citet{Glazunova2008} detected the 
signature of the tertiary component appears in the mean line profile of SZ~Psc. More recently, using the similar method, \citet{Xiang2016} have also confirmed the existence of the 
tertiary star in the SZ~Psc system and obtained its mass and orbital period of about 0.9~M$_{\sun}$ and 1283$\pm$10~days, respectively. The contribution of the tertiary star to the 
luminosity of the system was also estimated to be about 5\%, which is in good agreement with the contribution of about 3--4\% derived by \citet{Eaton2007}.\\
\indent
SZ~Psc belongs to the RS~CVn class of variable stars \citep{Hall1976, Fekel1986}, which are characterized by particularly intense magnetic activity that manifests itself in the form of
remarkable photometric variability caused by changing spot coverage, chromospheric activity, transition region and coronal emission. The starspot activity on SZ~Psc has been studied 
through photometry by several authors \citep[e.g.][]{Eaton1979, Lanza2001, Kang2003, Eaton2007}. The long-term starspot evolution, a possible activity cycle, and orbital period variation 
have been investigated by \citet{ Lanza2001}, based on an extended sequence of light curves obtained between 1957 and 1998. The first Doppler images of SZ~Psc derived by \citet{Xiang2016} 
show that the K1~IV star exhibits pronounced high-latitude spots as well as numerous intermediate- and low-latitude spot groups during the entire observing seasons. Moreover, the system 
exhibits a high level of chromospheric activity associated with the K1~IV primary component, as demonstrated by strong chromospheric emission in the $\mbox{Mg~{\sc ii}}$ h \& k, 
$\mbox{Ca~{\sc ii}}$ H \& K, H$_{\alpha}$, and $\mbox{Ca~{\sc ii}}$ IRT lines \citep{Jakate1976, Bopp1981, Ramsey1981, Huenemoerder1984, Fernandez1986, Popper1988, Frasca1994, 
Kang2003, Eaton2007, Zhang2008, Cao2012}. \citet{Zhang2008} analyzed several chromospheric activity indicators $\mbox{Ca~{\sc ii}}$ IRT, H$_{\alpha}$, $\mbox{Na~{\sc i}}$ D$_{1}$, 
D$_{2}$ doublet, and $\mbox{He~{\sc i}}$ D$_{3}$ lines by using the spectral subtraction technique and found that the chromospheric activity emissions of the $\mbox{Ca~{\sc ii}}$ 
$\lambda$8542, $\lambda$8662, and H$_{\alpha}$ lines are more strong near the two quadratures of the system.\\
\indent
The very unusual behavior of the H$_{\alpha}$ line of SZ~Psc has been reported several times. For example, \citet{Bopp1981} detected a remarkable H$_{\alpha}$ outburst in 1978, with the line 
profile evolving from weak absorption to a broad double-peaked emission during a few nights, and attributed this change to a transient mass-transfer event. In 1979, another H$_{\alpha}$ eruption 
event was observed by \citet{Ramsey1981} and interpreted as an intense flare-like event. \citet{Huenemoerder1984} also found a large H$_{\alpha}$ outburst during and after which the profiles 
suggest a circumstellar origin. Moreover, excess absorption features in the subtracted H$_{\alpha}$ profiles caused by prominence-like material have been discussed by \citet{Zhang2008} and 
\citet{Cao2012}.\\
\indent
X-ray emission from the SZ~Psc system was first detected by \citet{Walter1981} using the {\it Einstein Observatory} IPC, demonstrating that RS~CVn systems as a class are producers of copious 
soft X-ray emission. X-ray variability is quite common in such systems \citep{Fuhrmeister2003} and a strong X-ray flare-like event on the SZ~Psc system was detected by the Gas Slit Camera (GSC) 
of the Monitor of All-sky X-ray Image (MAXI) on 2011 November~5 \citep{Negoro2011}, which lasted for about five hours; we note that the occurrence of this event was very close to our observing 
run. Also, the X-ray luminosity of this flare-like event was two times larger than the previous one observed by the MAXI/GSC on 2009 December~4 \citep[also see][]{Negoro2011}.\\
\begin{table}
   \centering
   \caption{Observing log of SZ~Psc.}
   \tabcolsep 0.35cm
   \label{table1}
   \begin{tabular}{cccc}
   \hline
   \multicolumn{1}{c}{\bf{UT}} &
   \multicolumn{1}{c}{\bf{HJD}} &
   \multicolumn{1}{c}{\bf{Phase}} &
   \multicolumn{1}{c}{\bf{Exp.time}}\\
   \multicolumn{1}{c}{(hh:mm:ss)}&
   \multicolumn{1}{c}{(2,455,000+)}&
   \multicolumn{1}{c}{}&
   \multicolumn{1}{c}{(s)}\\
\hline
\multicolumn{4}{c}{\bf{October~24, 2011}}\\
12:32:22&859.0269&0.8760&600\\
12:51:43&859.0403&0.8794&900\\
13:09:34&859.0527&0.8825&900\\
13:26:36&859.0645&0.8855&900\\
13:43:35&859.0763&0.8885&900\\
14:00:50&859.0883&0.8915&900\\
14:15:31&859.0985&0.8941&600\\
14:27:36&859.1069&0.8962&600\\
14:39:36&859.1152&0.8983&600\\
14:51:37&859.1236&0.9004&600\\
15:03:40&859.1319&0.9025&600\\
15:15:42&859.1403&0.9046&600\\
15:27:51&859.1487&0.9068&600\\
15:39:55&859.1571&0.9089&600\\
15:51:58&859.1655&0.9110&600\\
16:04:01&859.1738&0.9131&600\\
16:16:02&859.1822&0.9152&600\\
16:28:03&859.1905&0.9173&600\\
16:40:05&859.1989&0.9194&600\\
16:52:08&859.2073&0.9215&600\\
17:04:09&859.2156&0.9236&600\\
17:16:18&859.2240&0.9258&600\\
17:28:26&859.2325&0.9279&600\\
17:40:30&859.2408&0.9300&600\\
17:52:35&859.2492&0.9321&600\\
18:04:36&859.2576&0.9342&600\\
18:16:40&859.2660&0.9363&600\\
\hline
\multicolumn{4}{c}{\bf{October~25, 2011}}\\
13:56:25&860.0852&0.1429&600\\
14:08:37&860.0936&0.1450&600\\
14:20:44&860.1020&0.1472&600\\
14:32:44&860.1104&0.1493&600\\
14:44:43&860.1187&0.1514&600\\
14:56:46&860.1271&0.1535&600\\
15:08:47&860.1354&0.1556&600\\
15:20:49&860.1439&0.1577&600\\
15:32:50&860.1521&0.1598&600\\
15:44:53&860.1605&0.1619&600\\
15:56:54&860.1688&0.1640&600\\
16:08:54&860.1772&0.1661&600\\
16:20:57&860.1855&0.1682&600\\
16:33:01&860.1939&0.1703&600\\
16:45:05&860.2023&0.1724&600\\
16:57:09&860.2107&0.1745&600\\
17:09:11&860.2190&0.1767&600\\
17:21:14&860.2274&0.1788&600\\
17:33:24&860.2358&0.1809&600\\
17:45:27&860.2442&0.1830&600\\
17:57:45&860.2528&0.1852&600\\
18:10:01&860.2613&0.1873&600\\
18:22:02&860.2696&0.1894&600\\
18:34:03&860.2780&0.1915&600\\
   \hline
   \end{tabular}
\end{table}
\indent
In summary, SZ~Psc is a notable star showing strong magnetic activity with very variable activity features. In this paper, we present the results of our time-resolved spectroscopic 
observations and simultaneous analysis of several optical chromospheric activity indicators. In Section~2, we provide the details of our spectroscopic observations and data reduction. 
The procedure of the spectral analysis and the behavior of chromospheric activity indicators are described in Section~3, and in Section~4, the magnetic activity phenomena observed 
on SZ~Psc are discussed in detail. Finally, we present the conclusions in Section~5.\\
\section{Spectroscopic observations and data reduction}
\begin{figure*}
      \centering
      \includegraphics[width=15cm,height=22cm]{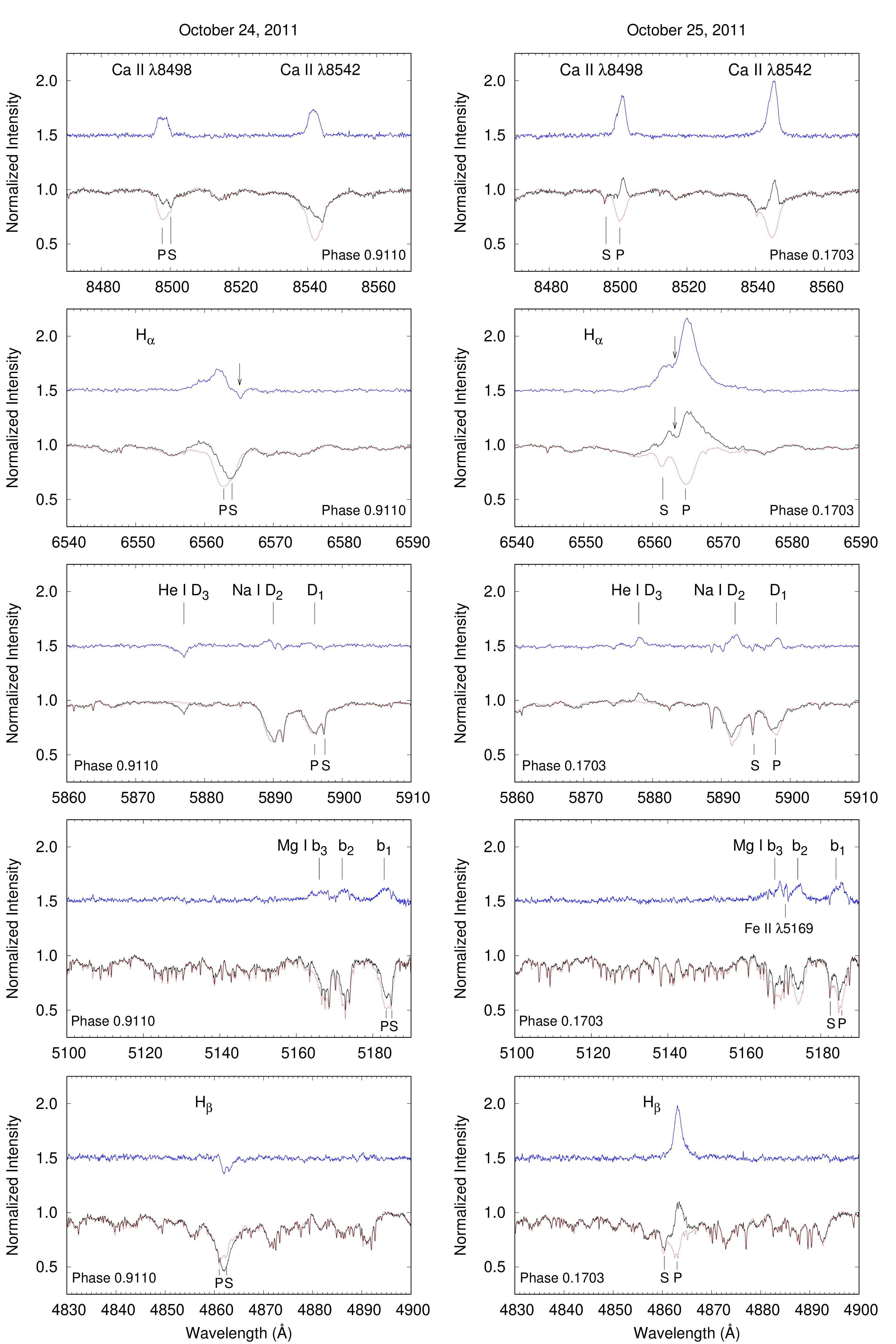}
      \caption{Examples of the observed, synthesized, and subtracted spectra for the $\mbox{Ca~{\sc ii}}$ $\lambda$8542 and 
      $\lambda$8498, H$_{\alpha}$, $\mbox{Na~{\sc i}}$ D$_{1}$, D$_{2}$ doublet, $\mbox{He~{\sc i}}$ D$_{3}$, 
      $\mbox{Mg~{\sc i}}$~b triplet, and H$_{\beta}$ line spectral regions. Left panels are examples of our observations on 
      October~24, while right panels show examples of October~25. For each panel, the lower solid-line is the observed 
      spectrum, the dotted line represents the synthesized spectrum and the upper spectrum is the subtracted one, shifted for better 
      display. ``P'' and ``S" indicate the primary and secondary components of the SZ~Psc system, respectively. Arrows indicate the 
      absorption features presented in the line profiles.}
      \label{figure1}
\end{figure*}

\indent
The time series of high-resolution spectra of SZ~Psc analyzed and discussed in this paper were obtained during two consecutive observing nights on October~24 and 25, 2011. Our 
observations were carried out with the direct echelle mode (DEM) of the Lijiang Exoplanet Tracker (LiJET) mounted on the 2.4-m telescope \citep{Fan1025} at the Lijiang station, 
administered by Yunnan Observatories, Chinese Academy of Sciences. The echelle spectrograph has a resolving power R~=~$\lambda$/$\Delta\lambda$~$\simeq$~28000 over 
the wavelength range from 3900 to 9500~\AA, and a $4096 \times 4096$-pixel CCD detector was used to record the spectra.\\
\indent
We provide an observing log of the SZ~Psc observations in Table~\ref{table1}, which includes the observing date, UT, the heliocentric Julian date (HJD), the orbital phase, and the exposure 
time. The orbital phases are calculated using the ephemeris
\begin{equation}
    HJD=2,449,284.4483+3^{d} .96566356\times E
\end{equation}
from \citet{Eaton2007}, where the epoch corresponds to the conjunction with the K1~IV star ``in front''. Our observations lasted for about 5.7 hours in the first night, while the 
observations had a duration of about 4.6 hours in the second night. In total, 51 spectra of SZ~Psc were obtained, with exposure times of typically 600 seconds and sometimes 900 seconds, 
providing a high time-resolution sampling for the system with a period of nearly four days. In addition, two inactive stars with spectral types and luminosity classes similar to the 
components of SZ~Psc to be used as templates for our spectral subtraction technique and a fast rotating early-type star to be used as a telluric template were also observed with the same 
instrumental setup.\\
\indent
The spectral reduction was performed with the IRAF$^{1}$\footnotetext[1]{IRAF is distributed by the National Optical Astronomy Observatories, which is operated by the Association 
of Universities for Research in Astronomy (AURA), Inc., under cooperative agreement with the National Science Foundation (NSF).} package following the standard procedures (image 
trimming, bias correction, scattered light subtraction, 1--D spectrum extraction, wavelength calibration, flat-field division, and continuum fitting). The wavelength calibration was obtained 
by using the emission lines of a Th-Ar lamp, and the flat-field division was performed by using the high signal-to-noise ratio (SNR) spectrum of the rapidly rotating early-type star HR~7894 
(B5~IV, $vsini$~=~330~km~s$^{-1}$). The spectrum of early-type star was normalized along the line profiles to provide not only a valuable flat-field template, but also was used to remove 
telluric absoprtion lines in the chromospheric activity line regions of interest with an interactive procedure in IRAF. Examples of removing the telluric lines in different spectral regions can 
be found in \citet{Gu2002}.\\
\indent
To compare the behavior of chromospheric activity indicators between our two observing nights in detail, we show some examples of the normalized $\mbox{Ca~{\sc ii}}$ $\lambda$8542, 
$\lambda$8498, H$_{\alpha}$, $\mbox{Na~{\sc i}}$ D$_{1}$, D$_{2}$ doublet, $\mbox{He~{\sc i}}$ D$_{3}$, $\mbox{Mg~{\sc i}}$~b triplet, and H$_{\beta}$ line profiles for each night 
in Fig.~\ref{figure1}. The observing time and orbital phases are also marked in the figure.\\
\section{Analysis of chromospheric activity indicators}
\indent
The wide wavelength coverage of our echelle spectra allows us to simultaneously study the behavior of all the optical chromospheric activity indicators formed at different atmospheric 
heights from the region of temperature minimum to the upper chromosphere, including the $\mbox{Ca~{\sc ii}}$ IRT ($\lambda$8662, $\lambda$8542 and $\lambda$8498), H$_{\alpha}$, 
H$_{\beta}$, $\mbox{Ca~{\sc ii}}$ H \& K, $\mbox{Na~{\sc i}}$ D$_{1}$, D$_{2}$ doublet, $\mbox{Mg~{\sc i}}$~b triplet ($\lambda$5183, $\lambda$5172 and $\lambda$5167), and 
$\mbox{He~{\sc i}}$ D$_{3}$ lines. Because the SNR is very low in the $\mbox{Ca~{\sc ii}}$ H \& K line regions and the $\mbox{Ca~{\sc ii}}$ $\lambda$8662 line is located at the very 
edge of the echelle frame, these three lines are not analyzed in the present paper.\\
\subsection{Spectral subtraction}
\indent
To extract the spectroscopic signatures caused by stellar activity from the observed spectrum, we apply the spectral subtraction technique to all observed spectra with the help of the 
program STARMOD \citep{Barden1985, Montes1995a, Montes1995b, Montes1997, Montes2000}. This subtraction technique has been widely used for chromospheric activity studies 
\citep[e.g.][]{Montes1995b, Montes1997, Montes2000, Gu2002, Frasca2008, Zhang2008, Cao2014, Cao2015, Cao2017, Zhang2016} and has also been applied for the detection of stellar 
prominences \citep{Hall1992, Gunn1997a, Gunn1997b, Cao2012}.\\
\indent
Although a spectral line from the third star has been found in the SZ~Psc spectrum, this third light contribution is very weak. Thus, spectra of two inactive stars, HR~7690 (K1~IV) and HR~7560 
(F8~V), are respectively used as templates for the primary and secondary component of the system to construct the synthesized spectra representing the non-active state of SZ~Psc. The rotational 
velocity ($v sin(i)$) value of each component of the SZ~Psc system is determined from our template spectra. Using the STARMOD program, we obtain the average $v sin(i)$ values of 78~km~s$^{-1}$ 
for the primary and 2~km~s$^{-1}$ for the secondary, using the high SNR spectra, spanning the wavelength regions 5920--6470~\AA~which contain many photospheric absorption lines. Our derived 
values are fully consistent with the results of 80 \& 5 km~s$^{-1}$ measured with different reference spectra by \citet{Eaton2007}, 78 \& 0 km~s$^{-1}$ by \citet{Zhang2008}, and 76.9 \& 3.4 km~s$^{-1}$ 
by \citet{Cao2012}. Moreover, the adopted intensity weight ratios are 0.83/0.17 for the $\mbox{Ca~{\sc ii}}$ $\lambda$8542, $\lambda$8498 spectral region, 0.815/0.185 for the H$_{\alpha}$ spectral 
region, 0.775/0.225 for the $\mbox{Na~{\sc i}}$ D$_{1}$, D$_{2}$ doublet and $\mbox{He~{\sc i}}$ D$_{3}$ spectral region, 0.725/0.275 for the $\mbox{Mg~{\sc i}}$~b triplet spectral region, and 
0.70/0.30 for the H$_{\beta}$ spectral region. Consequently, the synthesized spectra in all the observing phases are constructed by properly broadening and weighting the reference spectra to the 
values of $v sin(i)$ and the intensity weight ratios derived above, and shifting along the radial velocity axis. Finally, the subtracted spectra between the observed and the synthesized spectra are 
derived for SZ~Psc, which present activity contribution as excess emission or absorption features above or below the continuum level. We present some examples of our spectral subtraction in the 
$\mbox{Ca~{\sc ii}}$ $\lambda$8542, $\lambda$8498, H$_{\alpha}$, $\mbox{Na~{\sc i}}$ D$_{1}$, D$_{2}$ doublet, $\mbox{He~{\sc i}}$ D$_{3}$, $\mbox{Mg~{\sc i}}$~b triplet, and H$_{\beta}$ 
line regions in Fig.~\ref{figure1}.\\
\subsection{The behavior of chromospheric activity indicators}
\begin{figure*}
    \centering
     \includegraphics[width=9cm,height=14cm,angle=270]{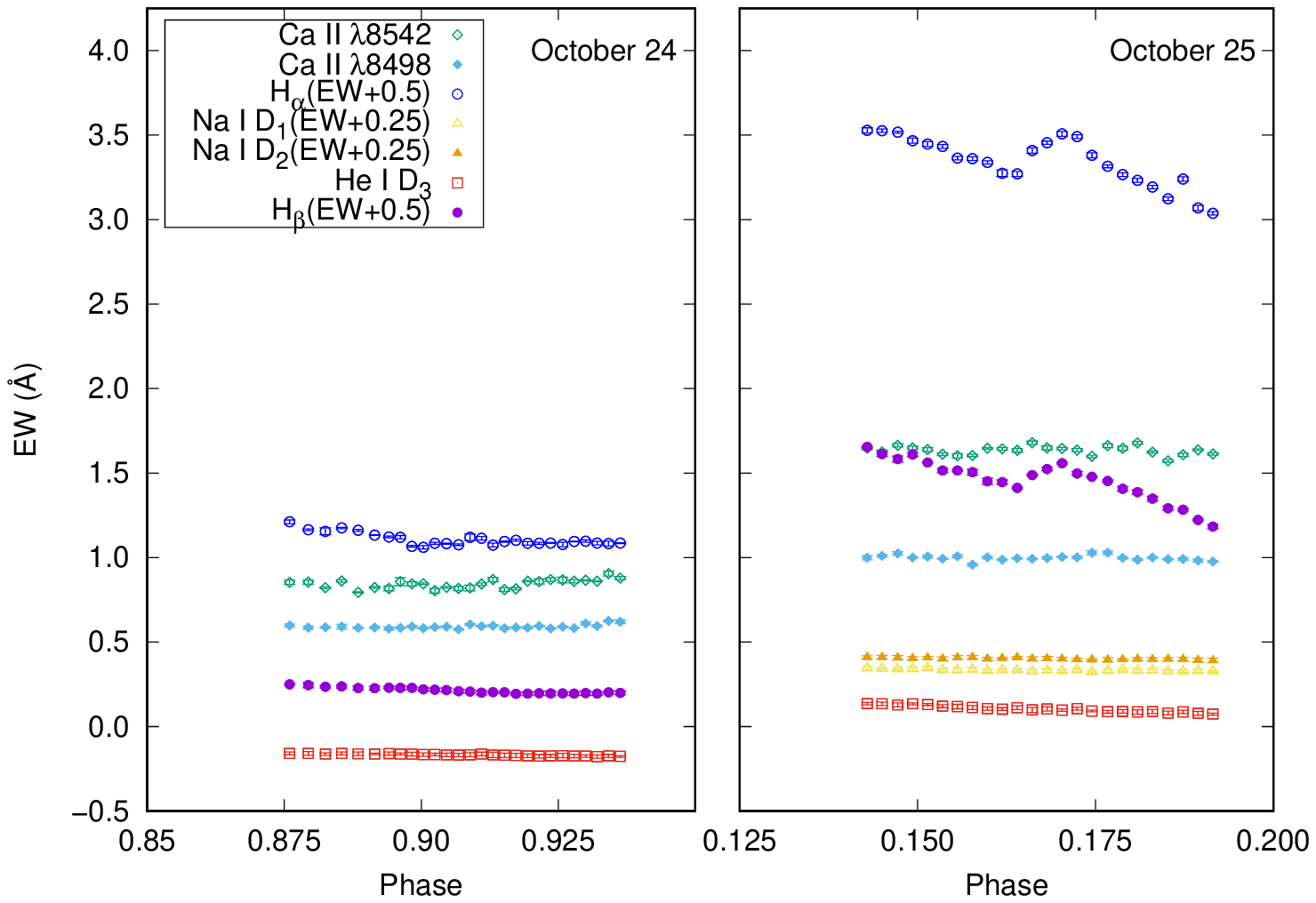}
     \caption{EWs of the subtracted $\mbox{Ca~{\sc ii}}$ $\lambda$8542, $\lambda$8498, H$_{\alpha}$, $\mbox{Na~{\sc i}}$ D$_{1}$, D$_{2}$
                    doublet, $\mbox{He~{\sc i}}$ D$_{3}$ and H$_{\beta}$ lines versus orbital phase during our two night observations. For better visibility, 
                    we shift EWs of the H$_{\alpha}$, $\mbox{Na~{\sc i}}$ D$_{1}$, D$_{2}$ doublet, and H$_{\beta}$ lines through adding offsets. The 
                    label identifying each optical chromospheric activity indicator is marked in the plot, and offset values are also given in the label.} 
      \label{figure2}
\end{figure*}

\begin{figure}
    \centering
     \includegraphics[width=6.5cm,height=8.5cm,angle=270]{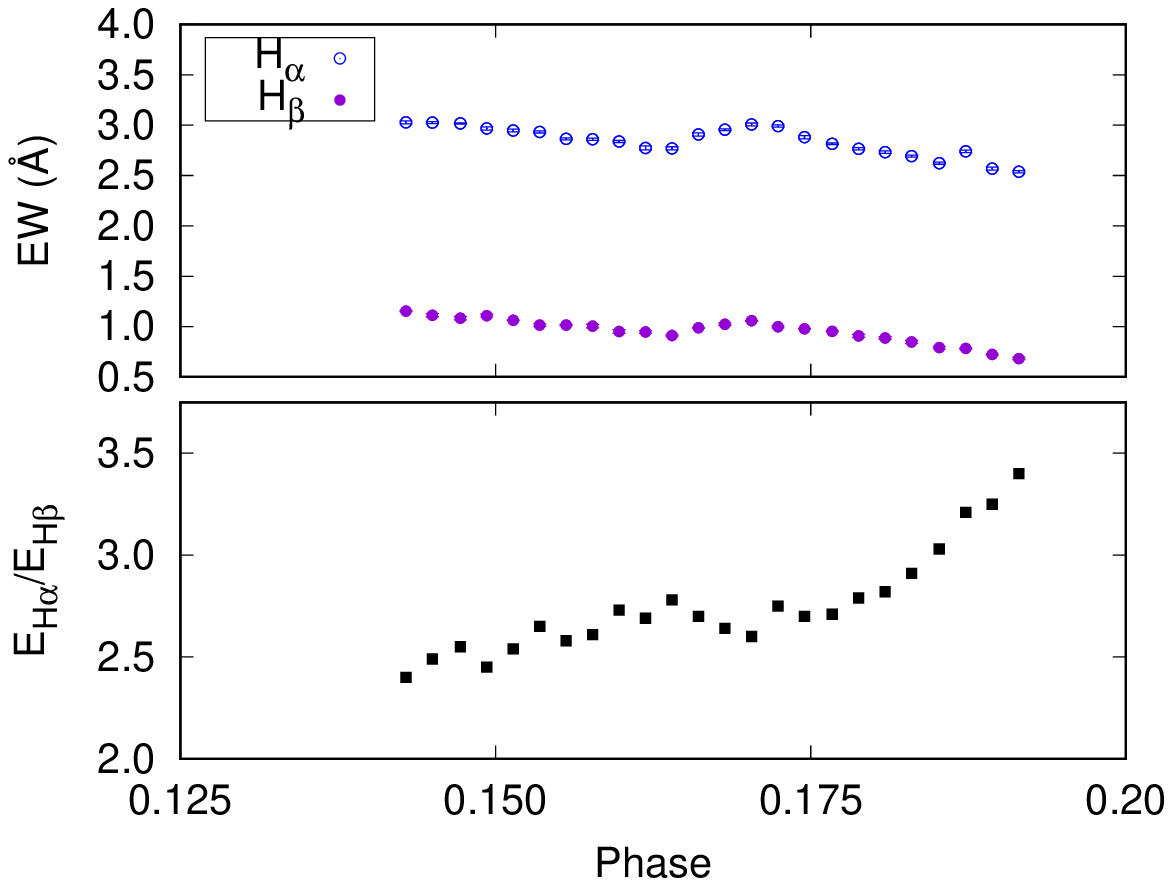}
     \caption{Comparison between the E$_{H{\alpha}}$/E$_{H{\beta}}$ ratios and the chromospheric emission variations in H$_{\alpha}$ and H$_{\beta}$ lines.} 
      \label{figure3}
\end{figure}

\begin{figure*}
    \centering
     \includegraphics[width=16cm,height=12.0cm]{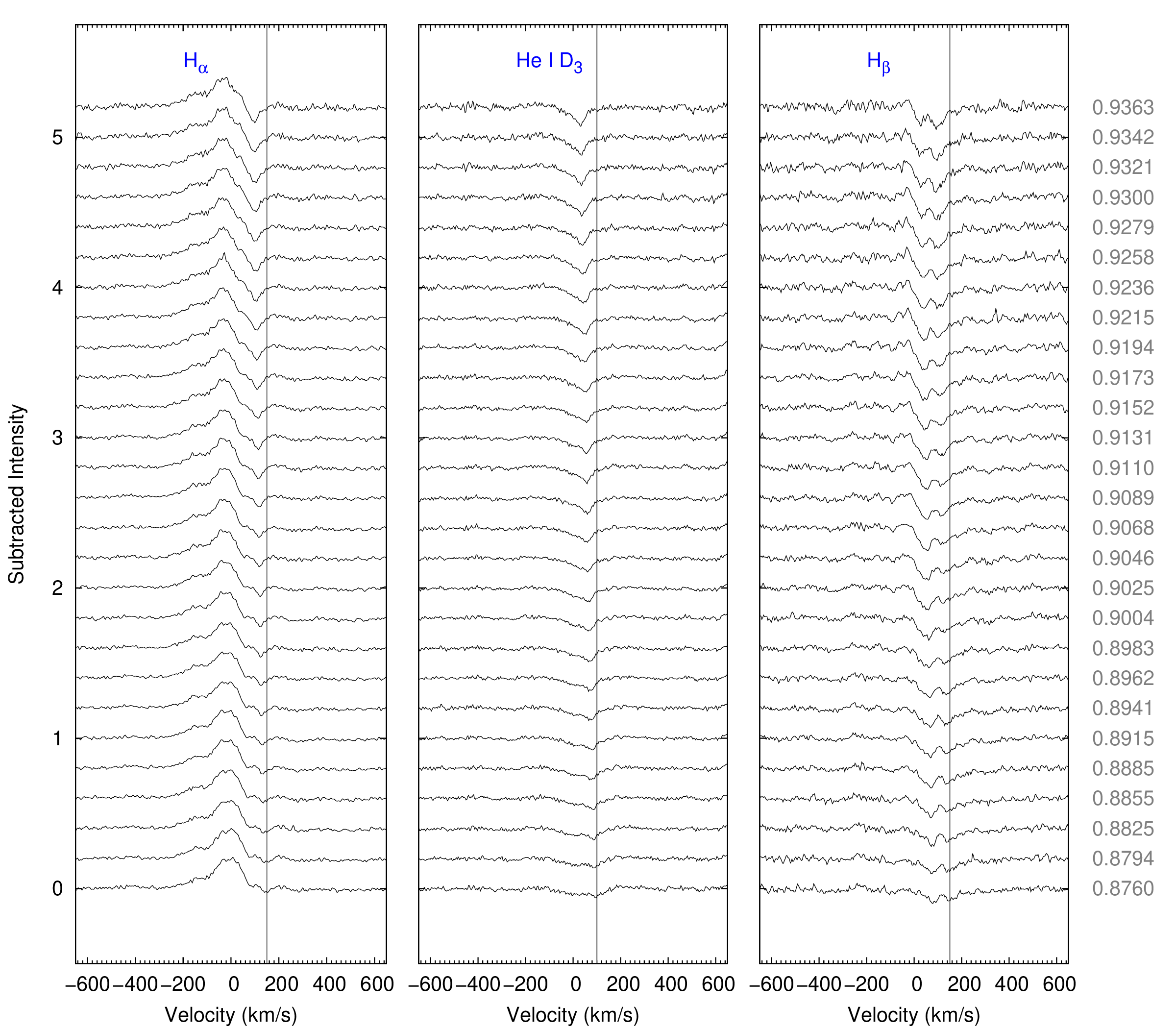}
     \caption{Series of the subtracted H$_{\alpha}$, $\mbox{He~{\sc i}}$ D$_{3}$ and H$_{\beta}$ line profiles obtained during the observations on October~24, 
                      and corrected to the rest velocity frame of the primary component of the SZ~Psc system. Spectra are shifted arbitrarily for better visibility. The vertical 
                      lines highlight the motion of absorption features and the phases are also marked in the plot.}
      \label{figure4}
\end{figure*}

\begin{figure*}
    \centering
    \includegraphics[width=13cm,height=12.0cm]{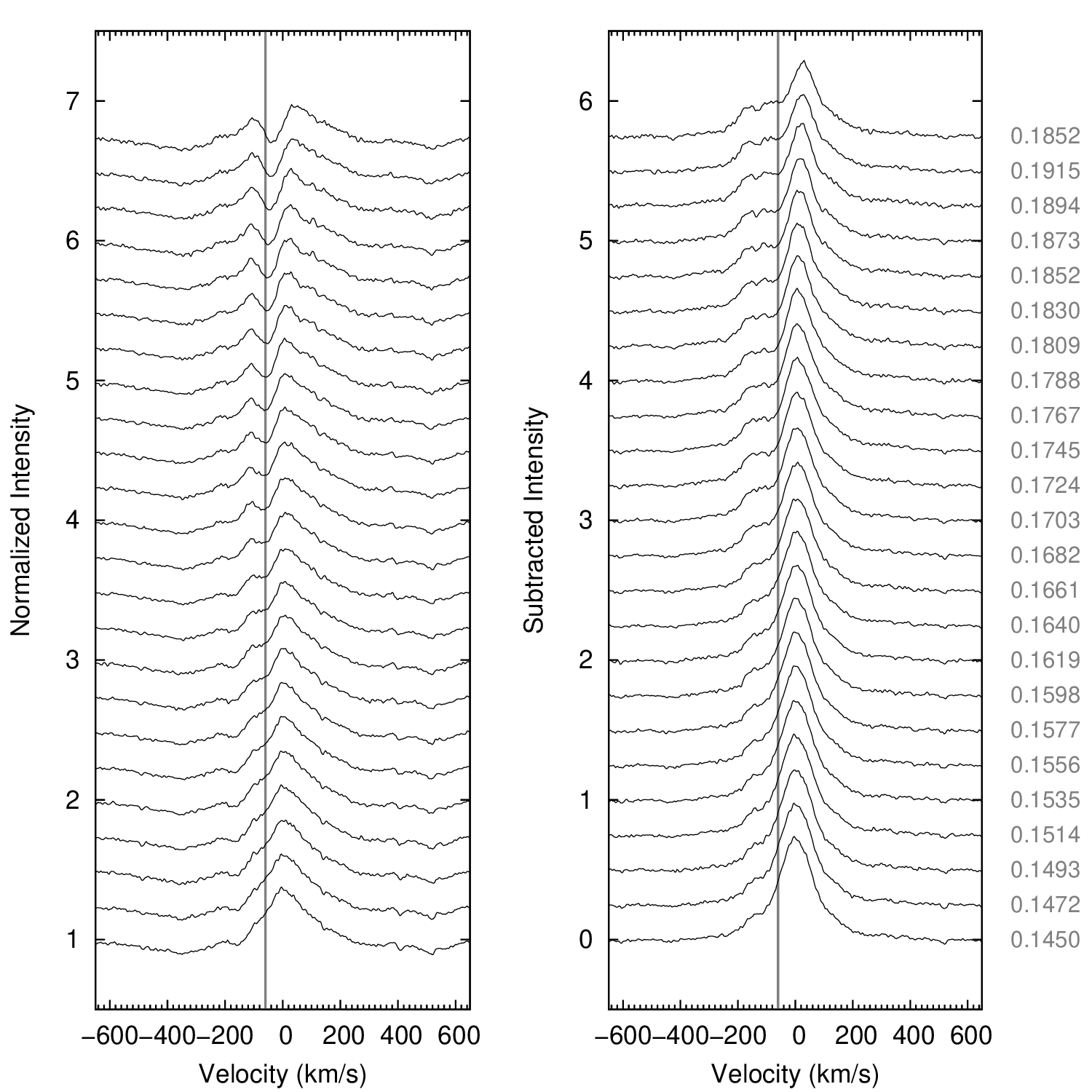}
    \caption{Same as Fig.~\ref{figure4}, but for the observed (left panel) and subtracted (right panel) H$_{\alpha}$ line profiles obtained during the observations 
                    on October~25.}
     \label{figure5}
\end{figure*}

\indent
After applying the spectral subtraction technique as shown in Fig.~\ref{figure1}, it becomes clear the excess emission is associated with the primary K1~IV component of the system, which dominates 
the activiy in the SZ~Psc system. The equivalent widths (EWs) of the subtracted $\mbox{Ca~{\sc ii}}$ $\lambda$8542, $\lambda$8498, H$_{\alpha}$, $\mbox{Na~{\sc i}}$ D$_{1}$, D$_{2}$ doublet, 
$\mbox{He~{\sc i}}$ D$_{3}$, and H$_{\beta}$ line profiles are measured with the splot task in IRAF, using the same method described in our previous papers \citep{Cao2015}, and are summarized 
in Table~\ref{table2} together with their errors; we also provide the ratios of the excess emission EW($\lambda$8542)/EW($\lambda$8498), and the E$_{H{\alpha}}$/E$_{H{\beta}}$ values for the 
observations on October~25, which are calculated from the excess emission EW(H$_{\alpha}$)/EW(H$_{\beta}$) ratios assuming that the continuum at H$_{\alpha}$ and H$_{\beta}$ line regions have 
a flux ratio appropriate to a blackbody (see the detailed description in Section~4.2). We plot the EWs of the subtraction profiles as a function of orbital phase in Fig.~\ref{figure2}, which shows that the 
level of activity is much higher in the second observing night; for example, the EWs of the subtracted H$_{\alpha}$ emission are about 3.5 $\sim$ 5 times stronger on October~25 than the previous night, 
while the EWs of the $\mbox{Ca~{\sc ii}}$ $\lambda$8542 and $\lambda$8498 lines are about 1.5 $\sim$ 2 times larger.\\
\indent
The EW($\lambda$8542)/EW($\lambda$8498) values are obtained around $\sim$1.4 in the first observing night, which are somewhat smaller than the values around $\sim$1.6 derived in the 
second night, when a strong optical flare decay was detected (also see the Section~4.2). These low ratios are consistent with the values found for several other stars with strong chromospheric 
activity \citep[e.g.][]{Montes2000, Gu2002, Lopez2003, Zhang2008, Galvez2009, Cao2014, Cao2015, Cao2017, Zhang2016}, which suggests that the $\mbox{Ca~{\sc ii}}$ IRT line emission arises 
from plage-like regions. The E$_{H{\alpha}}$/E$_{H{\beta}}$ ratios have also been usually used as a diagnostic for discriminating the presence of different structures on the stellar surface. 
As \citet{Huenemoerder1987} discussed, the low ratios in RS CVn-type stars are caused by plage-like regions, while prominence-like structures have high values. Similar results have also been 
reported by \citet{Hall1992} who found that low ratios ($\sim$ 1--2) can be achieved both in plages and prominences viewed against the disk, but high values ($\sim$ 3--15) can only be obtained 
in extended prominence-like structures viewed off the stellar limb. We obtain the ratios on SZ~Psc during the flare decay phase change from 2.40 to 3.40, which are not especially high and 
anti-correlated with the variation of activity emission shown in Fig.~\ref{figure3}, in which the EWs of H$_{\alpha}$ and H$_{\beta}$ line subtraction and the E$_{H{\alpha}}$/E$_{H{\beta}}$ 
values are plotted as a function of orbital phase. Especially when the EWs have an increasing oscillation during the gradually decrease, which is resulted from flare ejection (see the discussion in 
Section~4.2), the ratios have a similar anti-correlation feature. These facts suggest that the low E$_{H{\alpha}}$/E$_{H{\beta}}$ ratios and its variation might be dominantly associated with the 
flare decrease, and accompanied cool post-flare loops (see the Section~4.3) might play a part of role for the low ratios, which may have a state like absorbing prominence projected against the disk.\\
\indent
The chromospheric activity indicators $\mbox{Ca~{\sc ii}}$ $\lambda$8542, $\lambda$8498, H$_{\alpha}$, $\mbox{Na~{\sc i}}$ D$_{1}$, D$_{2}$ doublet, $\mbox{He~{\sc i}}$ D$_{3}$, 
$\mbox{Mg~{\sc i}}$~b triplet, and H$_{\beta}$ lines are characterized by deep absorption features in the observations on October~24. The subtracted H$_{\alpha}$ line shows obvious excess 
emission above the continuum level, but there is a local absorption feature appearing in the red wing of the subtracted profile (as indicated by the arrow in Fig.~\ref{figure1}). The H$_{\beta}$ 
line also exhibits an excess absorption in comparison to the inactive synthesized spectrum. An absorption feature also appears around the $\mbox{He~{\sc i}}$ D$_{3}$ line region in both observed 
and subtracted spectra. Moreover, similar absorption signatures are also detected in the other chromospheric activity indicators and the excess absorption features are very evident in the 
H$_{\alpha}$, H$_{\beta}$ and $\mbox{He~{\sc i}}$ D$_{3}$ lines. To further analyze the behavior of the excess absorption in our time series shown in Fig.~\ref{figure4}, we correct the subtracted 
line profiles to the rest frame of the primary star in the SZ~Psc system, and find that the velocity of the excess absorption features are gradually blue-shifted relative to the primary component and 
the intensity of the absorption strengthened with time during our observations.\\
\indent
In the observations on October~25, the $\mbox{Ca~{\sc ii}}$ $\lambda$8542, $\lambda$8498, H$_{\alpha}$, $\mbox{He~{\sc i}}$ D$_{3}$ and H$_{\beta}$ lines show strong emission features, 
which are remarkably different from our observations in the previous night. For the H$_{\alpha}$ line, an obvious absorption feature appears in the blue wing of the strong emission profile (as 
indicated by the arrow in Fig.~\ref{figure1}). Spectral subtraction shows that this absorption resulted in a strong depression in the blue wing of the subtracted emission profile. To clearly see the 
evolution of the line profile, we correct the observed and subtracted H$_{\alpha}$ spectra to the rest frame of the primary component along the velocity axis in Fig.~\ref{figure5}. Similar absorption 
features could also be found in the H$_{\beta}$ line, but they are much weaker than in the H$_{\alpha}$ line. The usual chromospheric flare diagnostic $\mbox{He~{\sc i}}$ D$_{3}$ line exhibits 
obvious emission above the continuum level, implying that a strong optical flare event took place on SZ~Psc. For the $\mbox{Ca~{\sc ii}}$ $\lambda$8542, $\lambda$8498 lines, the strong subtracted 
emission profiles are asymmetric due to the absorption feature appearing in the blue wing. Comparing with the behavior of the $\mbox{Na~{\sc i}}$ D$_{1}$, D$_{2}$ doublet and $\mbox{Mg~{\sc i}}$~b 
triplet lines obtained in the previous night, a strong fill-in of chromospheric emission from the primary component was observed in the subtracted spectra.\\
\begin{table*}
\centering
\begin{minipage}{170mm}
\caption{Measurements for the subtraction profiles of the $\mbox{Ca~{\sc ii}}$ $\lambda8542$, $\lambda8498$, H$_{\alpha}$, 
               $\mbox{Na~{\sc i}}$ D$_{1}$, D$_{2}$ doublet, $\mbox{He~{\sc i}}$~D$_{3}$ and H$_{\beta}$ lines, and 
               ratios of EW($\lambda$8542)/EW($\lambda$8498) and E$_{H{\alpha}}$/E$_{H{\beta}}$.}
\tabcolsep 0.125cm
\label{table2}
\begin{tabular}{@{}cccccccccc}
\hline
\bf{Phase} & \multicolumn{7}{c}{\bf{EW({\AA})}} & $\frac{\bf{EW(\lambda8542)}}{\bf{EW(\lambda8498)}}$& $\frac{\bf{E_{H{\alpha}}}}{\bf{E_{H{\beta}}}}$ \\
\cline{2-8}\\
&\bf{$\mbox{Ca~{\sc ii}}$ $\lambda8542$} & 
   \bf{$\mbox{Ca~{\sc ii}}$ $\lambda8498$} & 
   \bf{H$_{\alpha}$} & 
   \bf{$\mbox{Na~{\sc i}}$ D$_{1}$} &
   \bf{$\mbox{Na~{\sc i}}$ D$_{2}$} &
   \bf{$\mbox{He~{\sc i}}$ D$_{3}$} &
   \bf{H$_{\beta}$} & \\
\hline
&\multicolumn{7}{c}{\bf{October~24, 2011}}&&\\
\hline
0.8760& 0.853$\pm$0.013& 0.598$\pm$0.009& 0.711$\pm$0.011  &... &... &  -0.159$\pm$0.009& -0.250$\pm$0.005 & 1.43 & ...\\
0.8794& 0.854$\pm$0.012& 0.585$\pm$0.010& 0.664$\pm$0.005  &... &... &  -0.159$\pm$0.011& -0.255$\pm$0.015 & 1.46 & ...\\
0.8825& 0.821$\pm$0.005& 0.587$\pm$0.005& 0.655$\pm$0.020  &... &... &  -0.163$\pm$0.007& -0.265$\pm$0.007 & 1.40 & ...\\
0.8855& 0.861$\pm$0.008& 0.591$\pm$0.015& 0.675$\pm$0.004  &... &... &  -0.159$\pm$0.009& -0.263$\pm$0.009 & 1.46 & ...\\
0.8885& 0.794$\pm$0.003& 0.584$\pm$0.005& 0.661$\pm$0.006  &... &... &  -0.161$\pm$0.011& -0.272$\pm$0.012 & 1.36 & ...\\
0.8915& 0.823$\pm$0.005& 0.586$\pm$0.006& 0.633$\pm$0.002  &... &... &  -0.162$\pm$0.005& -0.273$\pm$0.011 & 1.40 & ...\\
0.8941& 0.817$\pm$0.014& 0.579$\pm$0.010& 0.622$\pm$0.005  &... &... &  -0.160$\pm$0.009& -0.270$\pm$0.006 & 1.41 & ...\\
0.8962& 0.858$\pm$0.023& 0.584$\pm$0.005& 0.620$\pm$0.010  &... &... &  -0.162$\pm$0.006& -0.271$\pm$0.012 & 1.47 & ...\\
0.8983& 0.844$\pm$0.013& 0.591$\pm$0.007& 0.566$\pm$0.005  &... &... &  -0.164$\pm$0.011& -0.271$\pm$0.010 & 1.43 & ...\\
0.9004& 0.845$\pm$0.005& 0.582$\pm$0.003& 0.560$\pm$0.010  &... &... &  -0.166$\pm$0.010& -0.280$\pm$0.011 & 1.45 & ...\\
0.9025& 0.804$\pm$0.014& 0.588$\pm$0.006& 0.584$\pm$0.006  &... &... &  -0.166$\pm$0.008& -0.282$\pm$0.009 & 1.37 & ...\\
0.9046& 0.824$\pm$0.005& 0.590$\pm$0.005& 0.582$\pm$0.003  &... &... &  -0.168$\pm$0.011& -0.284$\pm$0.013 & 1.40 & ...\\
0.9068& 0.817$\pm$0.013& 0.575$\pm$0.006& 0.576$\pm$0.005  &... &... &  -0.169$\pm$0.010& -0.291$\pm$0.012 & 1.42 & ...\\
0.9089& 0.820$\pm$0.017& 0.604$\pm$0.005& 0.620$\pm$0.019  &... &... &  -0.167$\pm$0.012& -0.293$\pm$0.009 & 1.36 & ...\\
0.9110& 0.843$\pm$0.007& 0.594$\pm$0.005& 0.614$\pm$0.011  &... &... &  -0.163$\pm$0.011& -0.300$\pm$0.013 & 1.42 & ...\\
0.9131& 0.870$\pm$0.013& 0.597$\pm$0.006& 0.574$\pm$0.012  &... &... &  -0.168$\pm$0.015& -0.297$\pm$0.006 & 1.46 & ...\\
0.9152& 0.810$\pm$0.010& 0.582$\pm$0.008& 0.595$\pm$0.004  &... &... &  -0.170$\pm$0.013& -0.298$\pm$0.007 & 1.39 & ...\\
0.9173& 0.815$\pm$0.006& 0.586$\pm$0.003& 0.602$\pm$0.006  &... &... &  -0.171$\pm$0.012& -0.306$\pm$0.010 & 1.39 & ...\\
0.9194& 0.860$\pm$0.004& 0.585$\pm$0.007& 0.584$\pm$0.010  &... &... &  -0.173$\pm$0.010& -0.305$\pm$0.009 & 1.47 & ...\\
0.9215& 0.858$\pm$0.015& 0.595$\pm$0.004& 0.584$\pm$0.006  &... &... &  -0.176$\pm$0.011& -0.303$\pm$0.011 & 1.44 & ...\\
0.9236& 0.870$\pm$0.011& 0.580$\pm$0.005& 0.586$\pm$0.003  &... &... &  -0.174$\pm$0.009& -0.304$\pm$0.014 & 1.50 & ...\\
0.9258& 0.867$\pm$0.017& 0.590$\pm$0.005& 0.577$\pm$0.013  &... &... &  -0.172$\pm$0.013& -0.304$\pm$0.009 & 1.47 & ...\\
0.9279& 0.860$\pm$0.010& 0.583$\pm$0.002& 0.595$\pm$0.002  &... &... &  -0.175$\pm$0.011& -0.305$\pm$0.011 & 1.48 & ...\\
0.9300& 0.865$\pm$0.003& 0.610$\pm$0.010& 0.597$\pm$0.007  &... &... &  -0.175$\pm$0.007& -0.302$\pm$0.013 & 1.42 & ...\\
0.9321& 0.860$\pm$0.004& 0.595$\pm$0.007& 0.585$\pm$0.010  &... &... &  -0.178$\pm$0.010& -0.305$\pm$0.010 & 1.45 & ...\\
0.9342& 0.905$\pm$0.014& 0.625$\pm$0.005& 0.582$\pm$0.014  &... &... &  -0.174$\pm$0.009& -0.298$\pm$0.009 & 1.45 & ...\\
0.9363& 0.878$\pm$0.009& 0.620$\pm$0.011& 0.585$\pm$0.003  &... &... &  -0.177$\pm$0.005& -0.301$\pm$0.011 & 1.42 & ...\\
\hline
&\multicolumn{7}{c}{\bf{October~25, 2011}}&&\\
\hline
0.1429& 1.648$\pm$0.009& 0.999$\pm$0.012& 3.028$\pm$0.015& 0.095$\pm$0.011& 0.159$\pm$0.009& 0.135$\pm$0.007& 1.154$\pm$0.005 & 1.65 & 2.40\\
0.1450& 1.623$\pm$0.015& 1.010$\pm$0.008& 3.026$\pm$0.010& 0.092$\pm$0.005& 0.159$\pm$0.011& 0.134$\pm$0.011& 1.113$\pm$0.015 & 1.61 & 2.49\\
0.1472& 1.664$\pm$0.010& 1.024$\pm$0.011& 3.017$\pm$0.005& 0.090$\pm$0.007& 0.156$\pm$0.013& 0.128$\pm$0.012& 1.083$\pm$0.017 & 1.63 & 2.55\\
0.1493& 1.649$\pm$0.011& 1.000$\pm$0.004& 2.967$\pm$0.015& 0.090$\pm$0.011& 0.153$\pm$0.010& 0.134$\pm$0.006& 1.109$\pm$0.013 & 1.65 & 2.45\\
0.1514& 1.639$\pm$0.012& 1.005$\pm$0.008& 2.947$\pm$0.014& 0.095$\pm$0.010& 0.156$\pm$0.009& 0.130$\pm$0.005& 1.062$\pm$0.002 & 1.63 & 2.54\\
0.1535& 1.611$\pm$0.008& 0.993$\pm$0.004& 2.933$\pm$0.011& 0.083$\pm$0.009& 0.151$\pm$0.011& 0.120$\pm$0.009& 1.015$\pm$0.013 & 1.62 & 2.65\\
0.1556& 1.602$\pm$0.013& 1.008$\pm$0.009& 2.864$\pm$0.010& 0.084$\pm$0.013& 0.157$\pm$0.012& 0.118$\pm$0.013& 1.015$\pm$0.006 & 1.59 & 2.58\\
0.1577& 1.604$\pm$0.007& 0.957$\pm$0.007& 2.860$\pm$0.011& 0.087$\pm$0.012& 0.159$\pm$0.010& 0.112$\pm$0.011& 1.005$\pm$0.017 & 1.68 & 2.61\\
0.1598& 1.646$\pm$0.005& 1.001$\pm$0.007& 2.838$\pm$0.010& 0.080$\pm$0.010& 0.151$\pm$0.009& 0.106$\pm$0.010& 0.951$\pm$0.019 & 1.64 & 2.73\\
0.1619& 1.643$\pm$0.009& 0.987$\pm$0.003& 2.774$\pm$0.020& 0.083$\pm$0.009& 0.154$\pm$0.005& 0.103$\pm$0.008& 0.946$\pm$0.013 & 1.66 & 2.69\\
0.1640& 1.634$\pm$0.011& 0.996$\pm$0.004& 2.770$\pm$0.015& 0.082$\pm$0.007& 0.158$\pm$0.007& 0.111$\pm$0.011& 0.912$\pm$0.006 & 1.64 & 2.78\\
0.1661& 1.679$\pm$0.010& 0.992$\pm$0.005& 2.908$\pm$0.016& 0.076$\pm$0.011& 0.151$\pm$0.011& 0.098$\pm$0.015& 0.988$\pm$0.005 & 1.69 & 2.70\\
0.1682& 1.649$\pm$0.012& 0.997$\pm$0.007& 2.955$\pm$0.009& 0.081$\pm$0.010& 0.152$\pm$0.012& 0.104$\pm$0.013& 1.024$\pm$0.010 & 1.65 & 2.64\\
0.1703& 1.645$\pm$0.007& 1.003$\pm$0.003& 3.007$\pm$0.014& 0.078$\pm$0.008& 0.150$\pm$0.010& 0.097$\pm$0.007& 1.058$\pm$0.005 & 1.64 & 2.60\\
0.1724& 1.636$\pm$0.009& 1.001$\pm$0.008& 2.991$\pm$0.011& 0.083$\pm$0.009& 0.146$\pm$0.008& 0.106$\pm$0.009& 0.998$\pm$0.010 & 1.63 & 2.75\\
0.1745& 1.599$\pm$0.007& 1.028$\pm$0.013& 2.880$\pm$0.015& 0.072$\pm$0.007& 0.145$\pm$0.005& 0.092$\pm$0.005& 0.978$\pm$0.003 & 1.56 & 2.70\\
0.1767& 1.662$\pm$0.011& 1.029$\pm$0.011& 2.815$\pm$0.008& 0.079$\pm$0.010& 0.145$\pm$0.013& 0.087$\pm$0.008& 0.953$\pm$0.006 & 1.62 & 2.71\\
0.1788& 1.646$\pm$0.013& 0.998$\pm$0.006& 2.766$\pm$0.011& 0.085$\pm$0.008& 0.146$\pm$0.006& 0.090$\pm$0.010& 0.907$\pm$0.014 & 1.65 & 2.79\\
0.1809& 1.678$\pm$0.009& 0.987$\pm$0.005& 2.732$\pm$0.012& 0.079$\pm$0.011& 0.149$\pm$0.009& 0.086$\pm$0.013& 0.887$\pm$0.013 & 1.70 & 2.82\\
0.1830& 1.624$\pm$0.006& 1.000$\pm$0.003& 2.693$\pm$0.010& 0.082$\pm$0.012& 0.146$\pm$0.010& 0.088$\pm$0.012& 0.848$\pm$0.015 & 1.62 & 2.91\\
0.1852& 1.572$\pm$0.008& 0.990$\pm$0.004& 2.623$\pm$0.010& 0.075$\pm$0.010& 0.148$\pm$0.008& 0.080$\pm$0.011& 0.792$\pm$0.012 & 1.59 & 3.03\\
0.1873& 1.608$\pm$0.010& 0.991$\pm$0.007& 2.740$\pm$0.013& 0.077$\pm$0.007& 0.149$\pm$0.011& 0.086$\pm$0.010& 0.783$\pm$0.007 & 1.62 & 3.21\\
0.1894& 1.637$\pm$0.005& 0.982$\pm$0.009& 2.569$\pm$0.015& 0.079$\pm$0.009& 0.143$\pm$0.012& 0.078$\pm$0.012& 0.723$\pm$0.005 & 1.67 & 3.25\\
0.1915& 1.613$\pm$0.007& 0.976$\pm$0.006& 2.537$\pm$0.010& 0.076$\pm$0.005& 0.141$\pm$0.013& 0.074$\pm$0.006& 0.683$\pm$0.012 & 1.65 & 3.40\\
\hline
\end{tabular}
\medskip{Note: Negative EWs indicate that the subtraction profiles characterized by absorption features.}
\end{minipage}
\end{table*}
\section{Discussion}
\subsection{Prominence activation}
\indent
In the observations on October~24, the $\mbox{He~{\sc i}}$ D$_{3}$ line shows an absorption feature and, further, there is unusual excess absorption appearing in the subtracted H$_{\alpha}$, 
H$_{\beta}$ and other lines. We interpret the absorption features as caused by a prominence while being seen in projection against a significant fraction of the stellar disk. Stellar prominences 
have been reported for the first time by \citet{Robinson1986} as transient absorption features passing through the rotationally broadened H$_{\alpha}$ profile of the rapidly rotating K0 dwarf 
star AB~Dor. The absorption features are thought to originate in cool clouds of mostly neutral material magnetically supported high above the stellar photosphere and forced to corotate with the 
star in a manner of reminiscent of solar quiescent prominences, which scatter the underlying chromospheric emission out of the line-of-sight as they transit the stellar disc \citep{Collier1989a, 
Collier1989b}. Stellar prominences with heights of several stellar radii lying at or beyond the corotation radius of the star have lifetimes on the order of one week \citep{Collier1989a, Collier1989b}. 
Similar quiescent stellar prominence characteristics have now been detected on several rapidly rotating single active stars, such as AB~Dor \citep{Collier1989a, Collier1989b}, G dwarfs in the 
$\alpha$~Per cluster \citep{Collier1992}, BO~Mic \citep{Jeffries1993, Dunstone2006, Wolter2008}, HK~Aqr \citep{Byrne1996}, PZ~Tel \citep{Barnes2000}, RX~J1508.6-4423 \citep{Donati2000} and 
RE~1816+514 \citep{Eibe1998}. Moreover, \citet{Hall1992} also found prominence-like hints in eight of ten RS~CVn-type systems studied during their survey and concluded that stellar prominences 
may be common phenomena in this type stars.\\
\indent
Because the primary is very active in the SZ~Psc system, we infer that prominence is associated with this star. Although the absorption features appear very near the velocity of the secondary 
(see Fig.~\ref{figure1}), moreover, it is uncertain whether prominence is obscuring the part of the secondary. If this situation happened, the absorption features caused by prominence which is 
associated with the primary would has redward motion relative to the velocity of the secondary due to the orbital motion of the system during our observations. But the fact is that 
there is no obvious shift for the absorption features when we correct the subtracted line profiles to the rest frame of the secondary, unless prominence has motion of its own which result in the 
absorption shift and just exactly offset the redward motion. However, taking into account the line profile of the primary is very broad and thus it is more possible the absorption features are 
resulted from prominence which is absorbing the emission and continuum of the red wing from the primary, especially at the end of the observations on October~24 in the H$_{\alpha}$ line, 
we tend to think that prominence is obscuring the part of the primary.\\
\indent
According to the gradual blue-shift of the absorption relative to the underlying spectrum of the primary star and the increasing strength, this absorption could be understood in terms of an 
accelerated upward motion of the prominence material. As observed for the Sun, preceding the onset of a large (typically two-ribbon) flare, there might be transient signatures of flare-related 
activity occurring in the activity region itself or its neighborhood. It is widely accepted that prominence (filament) activation is closely tied to two-ribbon flares \citep{Ding2003, Sterling2005, 
Chifor2007}. Prior to the flare, a prominence (filament), located at the neutral lines of a two-ribbon flare, will become unstable due to some magnetic processes \citep[e.g.][]{Mikic1994, Forbes1995, Chen2000}, 
appearing as a rising motion or even an eruption. For our observations, there is a strong optical flare event in the second observing night (see Section~4.2). More importantly, both the excess 
absorption event and the optical flare observed during two consecutive nights may have occurred at the same hemisphere for a nearly four days periodic system, and therefore we think it is 
possible that both of them arise from the related active region. Furthermore, as discussed in next section, a X-ray flare-like event was detected at the same hemisphere after three orbital cycles 
and around the phase near our observations, indicating that there might be a strong activity region in this hemisphere and all of these activity phenomena are possibly associated with this region. 
In analogy to the activity phenomena observed in the solar case, we are led to infer that the excess absorption features would be produced due to a prominence activation event rising in height 
before the eruption of a strong optical flare.\\
\begin{figure}
    \centering
     \includegraphics[width=8.5cm,height=5.75cm]{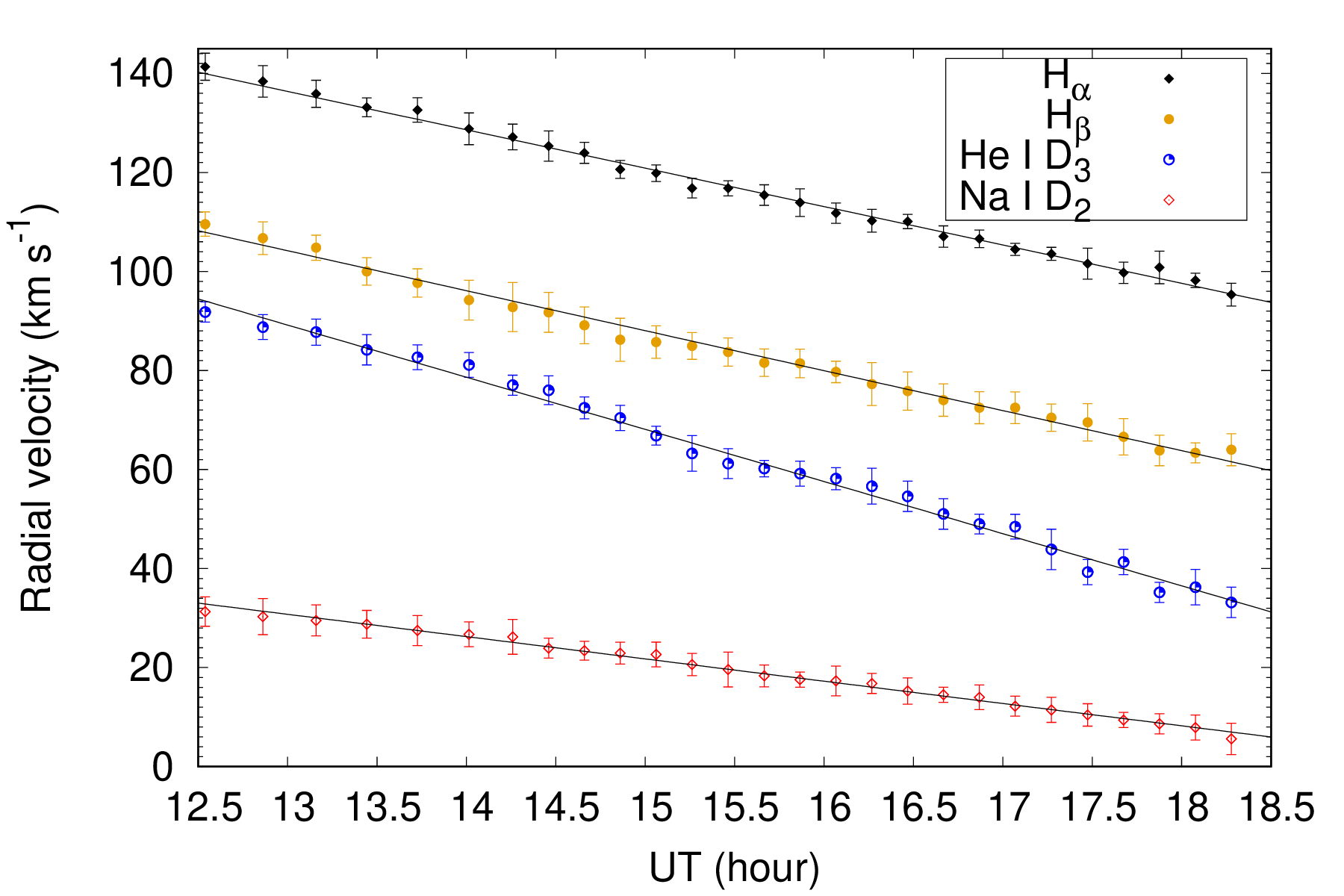}
     \caption{Radial velocities of excess absorption relative to the primary component of the system in the H$_{\alpha}$, 
                    $\mbox{Na~{\sc i}}$ D$_{2}$, $\mbox{He~{\sc i}}$ D$_{3}$ and H$_{\beta}$ lines during the 
                    observations on October 24. Solid lines represent the linear fits, and the label identifying each line 
                    is also marked in the plot.}
      \label{figure6}
\end{figure}
\indent
We measure the radial velocities of the excess absorption features relative to the primary component of the system in the chromospheric activity indicators H$_{\alpha}$, $\mbox{Na~{\sc i}}$ 
D$_{2}$, $\mbox{He~{\sc i}}$ D$_{3}$ and H$_{\beta}$ lines in the observations on October~24, and plot them against the observing time in Fig.~\ref{figure6}. The measured radial velocities 
are very different in these chromospheric lines, probably due to the fact that they are formed at different atmospheric heights, which lets the absorption features more or less linearly drift in 
RV with different rates. Using the measured drift rates, we can make a rough estimate of the height of the prominence raised along the line-of-sight during observations on October~24 and find 
values of 0.4~R$_{\sun}$ for the $\mbox{Na~{\sc i}}$ D$_{2}$ line formed at the upper photosphere and lower chromosphere, 0.6~R$_{\sun}$ for the H$_{\alpha}$ and H$_{\beta}$ lines 
formed at the middle chromosphere, and 0.9~R$_{\sun}$ for the $\mbox{He~{\sc i}}$ D$_{3}$ line formed at the upper chromosphere.\\ 
\subsection{Optical flare}
\indent
Unlike the absorption feature encountered in the previous night, the chromospheric flare indicator $\mbox{He~{\sc i}}$ D$_{3}$ line shows obvious emission above the continuum level 
throughout the observations on October~25, which provides strong support for the occurrence of a large optical flare due to its high excitation potential \citep{Zirin1988}. Moreover, the 
observed stronger emission in the $\mbox{Ca~{\sc ii}}$ $\lambda$8542, $\lambda$8498, H$_{\alpha}$ and H$_{\beta}$ lines also confirms the event as flare activity. Furthermore, as 
marked in Fig.~\ref{figure1}, the $\mbox{Fe~{\sc ii}}~\lambda5169$ line near the $\mbox{Mg~{\sc i}}$~b~$\lambda5167$ line was also found in emission during the flare event, similar 
to what has also been observed in the young active K2 dwarf stars PW~And by \citet{Lopez2003} and LQ~Hya by \citet{Montes1999}.\\
\indent
Stellar flares share many features with solar flares.  Solar flares are powerful and explosive phenomena in the outer solar atmosphere, which typically include three basic phases during 
their evolution: the fast rise of the emission intensity, the maximum, and the gradual decay phase \citep{Schrijver2000}. For SZ~Psc, we did not capture the whole flare evolution from 
the initial outburst to the very end due to limited observations and the observing gap between our two observing nights. The overall trend of the flare intensity shows a decrease (see 
Table~\ref{table2} and Fig.~\ref{figure2}), especially for the H$_{\alpha}$ and H$_{\beta}$ lines, which suggests that our observation was part of the gradual decay phase of a flare 
and that the onset of this flare likely occurred around the gap between our two observing nights. Assuming that the first observation (at phase 0.1429) of October~25 corresponds to the 
flare maximum, we can derive a lower limit to the flare energy. We compute the stellar continuum flux $F_{H_{\alpha}}$~(erg cm$^{-2}$ s$^{-1}$ \AA$^{-1}$) in the H$_{\alpha}$ line 
region as a function of the color index $B-V$ ($\sim$ 0.846 for SZ~Psc; \citealt{Messina2008}) based on the calibration
\begin{eqnarray}
\log{F_{H_{\alpha}}}=[7.538-1.081(B-V)]\pm{0.33}\nonumber \\
0.0~\leq~B-V~\leq~1.4
\end{eqnarray}
by \citet{Hall1996}, and convert the EWs into the absolute surface fluxes $F_{S}$~(erg~cm$^{-2}$~s$^{-1}$). Following the method adopted by \citet{Montes1999} and \citet{Garcia2003}, 
we obtain the stellar continuum fluxes by using $F_{H_{\alpha}}$ corrected for the flux ratios $F_{6563}$/$F_{\lambda}$ for the other chromospheric activity lines.\\
\indent
The flux ratios were given by assuming a blackbody with contribution of both the primary and secondary component of the SZ~Psc system at the effective temperatures $T_{eff}$~=~4910~K 
and $T_{eff}$~=~6090~K \citep{Lanza2001}, in which the contribution ratios of two components have been obtained in the spectral subtraction technique. Therefore, the flare energies 
$L$~(erg~s$^{-1}$) in these lines were derived through converting the absolute surface fluxes into luminosities by using the radius R$_{\ast}$~=~6.0~R$_{\sun}$ of the K1~IV primary component 
\citep{Eaton2007}. We assumed that the optical flare occurred on the primary star, and also corrected the EWs to the total continuum before converting to absolute surface fluxes. The absolute 
surface fluxes and luminosities in the chromospheric activity lines at phase 0.1429 are listed in Table~\ref{table3}. The values for the energy released in the H$_{\alpha}$ line are of similar order 
of magnitude as the strong flares of other very active RS~CVn-type stars, such as V711~Tau \citep{Garcia2003, Cao2015}, UX~Ari \citep{Montes1996, Gu2002, Cao2017} and HK~Lac \citep{catalano1994}.\\
\begin{table}
\centering
\begin{minipage}{80mm}
\caption{Absolute surface fluxes $Fs$ and flare luminosities $L$ released by the chromospheric activity lines.}
\tabcolsep 0.28cm
\label{table3}
\begin{tabular}{@{}lcc}
\hline
\bf{Line}&     \bf{Fs}                           &                 \bf{L}      \\
       & ($\times~10^{7}$~erg~cm$^{-2}$~s$^{-1}$) & ($\times~10^{31}$~erg~s$^{-1}$) \\ 
\hline
$\mbox{Ca~{\sc ii}}~\lambda8542$ & 0.61 & 1.32\\
$\mbox{Ca~{\sc ii}}~\lambda8498$ & 0.37 & 0.81\\
H$_{\alpha}$                                          & 1.56 & 3.41\\
$\mbox{Na~{\sc i}}$~D$_{1}$           & 0.06 & 0.12\\
$\mbox{Na~{\sc i}}$~D$_{2}$           & 0.09 & 0.20\\
$\mbox{He~{\sc i}}$~D$_{3}$           & 0.08 & 0.17\\
H$_{\beta}$                                            & 0.76 & 1.65\\
\hline
\end{tabular}
\end{minipage}
\end{table}
\begin{figure*}
    \centering
    \includegraphics[width=17.0cm,height=8.0cm]{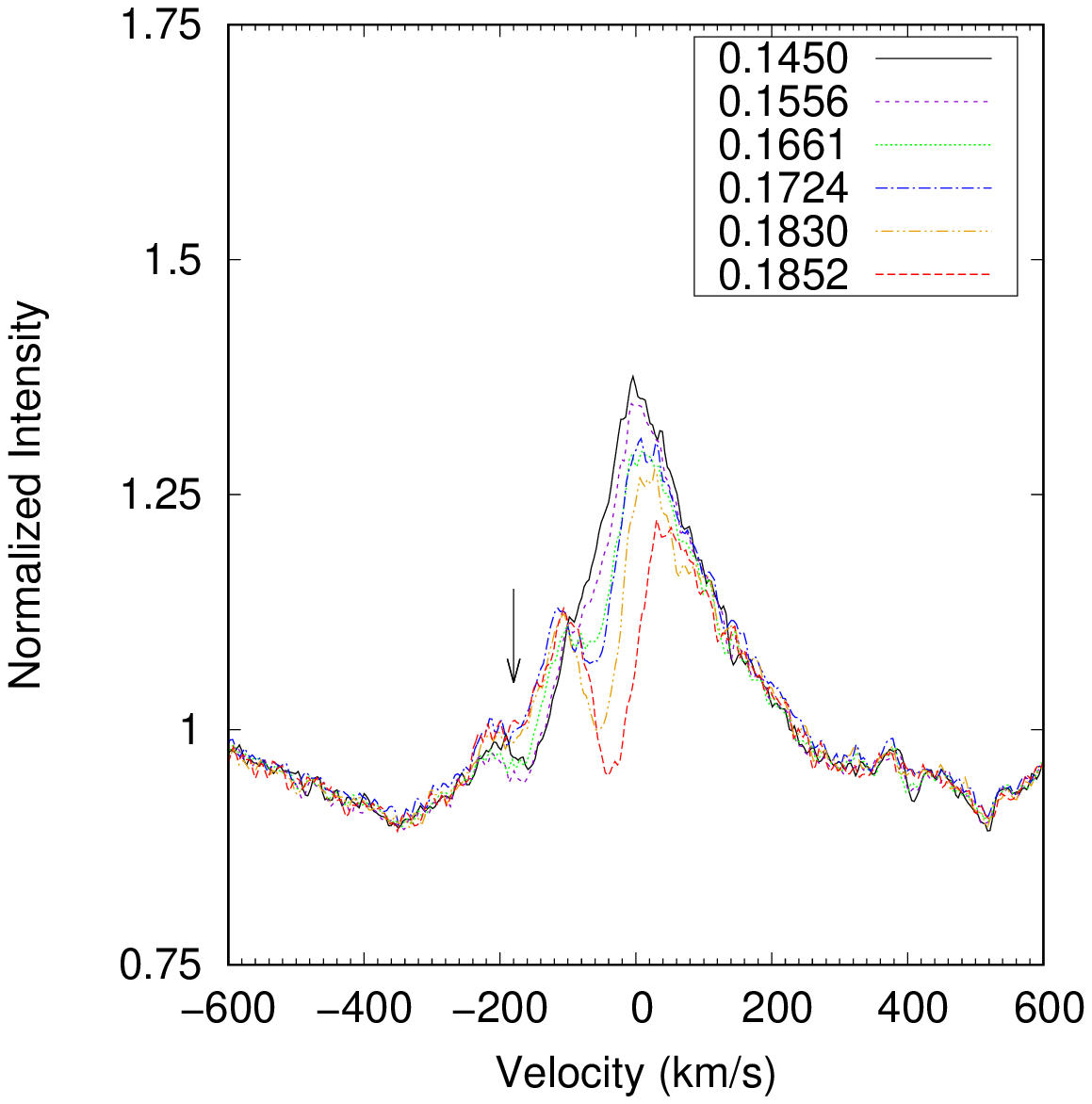}
    \caption{Same as Fig.~\ref{figure5}, but for the spectra selected at six phases.}
     \label{figure7}
\end{figure*}
\indent
Moreover, by comparison, we find that the X-ray flare-like event detected by the MAXI/GSC from 18:20~UT (= MJD~55870.764) to 23:03~UT (= MJD~55870.960) on 2011 November~5 
\citep{Negoro2011} occurred during phases 0.96--0.01, which is comparable to the phase between our two observing nights but three orbital cycles apart. Therefore, we argue that both 
X-ray flare-like event and our optical flare may have taken place at a same and long-lived active region over the surface of SZ~Psc.\\
\indent
Finally, in Fig.~\ref{figure7} we plot some of the observed and subtracted H$_{\alpha}$ spectra selected at six phases to illustrate the evolution of the line profile; it is clear that there is the 
notable enhancement in the far blue wing of the H$_{\alpha}$ line emission profiles at velocities between -150 and -200~km~s$^{-1}$ during our observations (as indicated by the arrow), and 
therefore an increasing oscillation occurred when the EWs gradually decrease during the flare decay phase (see Fig.~\ref{figure2}), similar to the finding derived by \citet{Eibe1999} for the 
late-type fast rotator BD+22$\degr$4409. The excess emission was explained by high velocity ejections of hot chromospheric material during the flare event; similar features could also be found 
in the H$_{\beta}$ line.\\
\subsection{Post-flare loops}
\indent 
We detected a fast evolved absorption feature in the blue wing of emission profile in the observations on October~25 (especially for the H$_{\alpha}$ line, see Fig.~\ref{figure5}) during 
the gradual decay phase of the flare. It could be explained by cool post-flare loops seen in projection against the bright flare background. Post-flare loops are arcade-like, developed systems, 
which are usually found during the gradual decay phase of large two-ribbon flares in the Sun and can last for several hours \citep{Sturrock1968, Kopp1976, Forbes1986}. Except for prominence 
activation event detected in our previous night observations, this post-flare loops also suggest our optical flare to be a two-ribbon flare.\\
\indent
As shown in Fig.~\ref{figure5}, the velocity of the absorption feature shows no distict change relative to the primary component of the system in the initial stage, but the strength gradually 
increases during our observations, it could be understood that the loops were rising their height at a nearly steady velocity of about -60~km~s$^{-1}$. Similar absorption feature 
at velocity -50~km~s$^{-1}$ in the H$_{\alpha}$ line profile during the decay of the flare had been observed by \citet{Eibe1999} on BD+22$\degr$4409 and also attributed to the 
development of cool dark flare loops that are seen against the flare background. Considering solar studies, the upward motion of the post-flare loops are caused by the systematic 
ascending of the magnetic reconnection site in the corona \citep{Sturrock1968, Kopp1976, Forbes1986}. For our observation, on the other hand, the absorption feature seems to show 
an obvious shift toward the red direction from phase 0.1809, which possibly suggests that the post-flare loops began to rise with a gradually decreasing velocity, consistent with what 
one typically observes in the solar case.\\
\section{Conclusions}
\indent
Our analysis of the  time-resolved spectroscopic observations of several chromospheric activity indicators (including the $\mbox{Ca~{\sc ii}}$ $\lambda$8542 and $\lambda$8498, 
H$_{\alpha}$, $\mbox{Na~{\sc i}}$ D$_{1}$, D$_{2}$, $\mbox{He~{\sc i}}$ D$_{3}$, $\mbox{Mg~{\sc i}}$~b triplet, and H$_{\beta}$ lines), taken during two consecutive observing 
nights on October 24 and 25, 2011, demonstrates the clear detection of a series of magnetic activity phenomena on the very active RS CVn-type star SZ~Psc, including a likely prominence 
activation event, an optical flare, and post-flare loops. The prominence activation occurred on SZ~Psc before the occurrence of a strong optical flare was observed in the observations on 
October~24, and we argue that this prominence event is possibly associated with the subsequent strong optical flare. The gradual decay phase of a large optical flare was detected in the 
observations on October~25, during which cool post-flare loops were seen as developing absorption features against bright flare background in the chromospheric lines, especially in the 
H$_{\alpha}$ line. Moreover, the prominence activation event and post-flare loops both indicate that optical flare on SZ~Psc can be classified as a two-ribbon flare.\\ 
\indent
To our knowledge, the detection of such a series of possibly associated magnetic activity phenomena in short period of time is rare in the stellar case. In order to further confirm 
this series of activity phenomena and investigate the detailed connection among them for the SZ~Psc system, we require more high-resolution spectral observations with higher 
temporal cadence.\\
\section*{Acknowledgements}
We are hugely grateful to those who contributed to the LiJET project. We would like to thank all the staff of the 2.4-m telescope of Lijiang station of Yunnan Observatories for their help and 
support during our observations. Funding for the telescope has been provided by the Chinese Academy of Sciences (CAS) and the People's Government of Yunnan Province. We acknowledge 
the support from the Project Based Personnel Exchange Program (PPP) with China Scholarship Council (CSC) and German Academic Exchange Service (DAAD) ([2016] 6041). We also thank the 
anonymous referee for helpful comments and suggestions, which led to significantly improvement in our manuscript. This work was financially supported by the National Natural Science 
Foundation of China (NSFC) under grant Nos. 10373023, 10773027 and 11333006, and the CAS through project No. KJCX2-YW-T24, and the CAS ``Light of West China'' Program.

\bsp	
\label{lastpage}
\end{document}